\documentclass{ifacconf}

\usepackage{graphicx}
\usepackage{natbib}
\usepackage{amsmath,amssymb,amsfonts}
\usepackage{algorithm}
\usepackage{algpseudocode}
\usepackage{bm}
\usepackage{booktabs}
\usepackage{mathrsfs}

\newcommand{\R}{\mathbb{R}}

\newcommand{\calA}{\mathcal{A}}
\newcommand{\calL}{\mathcal{L}}

\newcommand{\calS}{\mathcal{S}}

\newcommand{\diag}{\text{diag}}
\newcommand{\argmin}{\mathop{\mathrm{argmin}}}

\begin{document}
\begin{frontmatter}

\title{RDS-DeePC: Robust Data Selection for Data-Enabled Predictive
  Control via Sensitivity Score}

\author[Austin]{Jiachen Li}
\author[Austin]{Shihao Li}
\author[Austin]{Jian Chu}
\author[Austin]{Dongmei Chen}

\address[Austin]{The University of Texas at Austin,
  Austin, TX 78712 USA\\
  (e-mail: \{jiachenli, shihaoli01301, jian\_chu, dmchen\}@utexas.edu)}

\begin{abstract}
Data-Enabled Predictive Control (DeePC) is an established model-free
approach to predictive control, but it faces two open challenges:
computational complexity that scales cubically with dataset size and
performance degradation when data are corrupted. This
paper introduces Robust Data Selection DeePC (RDS-DeePC), a framework
that addresses both obstacles through influence function analysis. We
derive a sensitivity score quantifying the leverage each trajectory
segment exerts on the optimization solution and prove that
high-sensitivity segments correspond to outliers while low-sensitivity
segments represent consistent data. Selecting low-sensitivity segments
thus yields both computational efficiency and automatic outlier
filtering—without requiring data quality labels. For nonlinear
systems, we extend the framework via a two-stage online selection
approach accelerated by the LiSSA algorithm. Experiments on four
systems of increasing complexity—a DC motor (LTI), an inverted
pendulum, a planar quadrotor UAV tracking a figure-8 trajectory, and
a kinematic bicycle vehicle following a figure-8 path
—demonstrate that RDS-DeePC achieves 94–97\% clean data selection
and comparable or better tracking performance under 20\%
data corruption.
\end{abstract}


\end{frontmatter}

\section{Introduction}
\label{sec:intro}

Data-Enabled Predictive Control (DeePC) \citep{coulson2019data} is a
well-known framework for model-free predictive control. By leveraging
Willems' fundamental lemma \citep{willems2005note}, DeePC uses
input-output trajectory data directly to predict future system
behavior, bypassing explicit system identification. This behavioral
systems approach \citep{markovsky2021behavioral} has been applied to
power electronics \citep{huang2019data}, building climate control
\citep{lian2021adaptive}, autonomous vehicles \citep{elokda2021data},
and robotic systems \citep{salzmann2023neural}.

In practice, however, the deployment of DeePC is limited by two challenges
that this paper addresses.

\subsection{Challenge 1: Computational Intractability}

The computational cost of DeePC scales cubically with the number of
trajectory segments $T$ in the Hankel matrices. For a dataset of
$N_d$ samples and horizon parameters yielding $T = N_d - L + 1$
segments, the optimization at each Model Predictive Control (MPC)
step demands $O(T^3)$ operations with direct solvers, or $O(T^2)$
per iteration with iterative methods. This makes real-time control
difficult compared to fast
MPC implementations that exploit problem structure
\citep{wang2010fast}.

To illustrate, a dataset of $N_d = 5000$ samples yields $T \approx 4000$ segments; direct
solution then requires roughly $6 \times 10^{10}$ operations per MPC
step, making real-time control at typical sampling rates
(20--100~Hz) infeasible on standard hardware. As
\cite{luppi2024data} emphasize, data selection is not merely
beneficial but \emph{necessary} for online DeePC implementation.

\subsection{Challenge 2: Sensitivity to Data Quality}

Real-world trajectory data contain imperfections: sensor noise and
measurement errors, actuator faults producing input-output mismatch,
data drawn from different operating conditions or system
configurations, and corrupted or mislabeled recordings. When such
corrupted data enter the Hankel matrices, DeePC performance degrades,
sometimes to the point of failure.

Existing remedies include regularization
\citep{coulson2019regularized} and robust optimization
\citep{berberich2020data}. The learning-based MPC literature
\citep{hewing2020learning} has explored further strategies for
handling uncertain data. Although these methods improve nominal
robustness, none explicitly identify and remove problematic data
segments.

\subsection{Our Contribution: Solving Both Challenges at Once}

We introduce Robust Data Selection DeePC (RDS-DeePC), which uses
influence functions from robust statistics
\citep{hampel1974influence,cook1980characterizations} to derive a
\emph{sensitivity score} measuring the leverage each trajectory
segment exerts on the DeePC optimization. The key idea is that both
challenges can be addressed through a single mechanism:
high-sensitivity segments are high-leverage points that shift the
optimization solution, and when data quality is mixed, these
high-leverage points typically coincide with outliers or corrupted
recordings.

The contributions are threefold. First, we derive the sensitivity
score $\calS(z_j) = 2\lambda_g g_j^* p_j$ from influence function
theory \citep{koh2017understanding} and establish its interpretation
as a leverage measure. Second, we develop the RDS-DeePC algorithm for
LTI systems, which achieves robust control even when a large fraction
of data is corrupted, while also providing a large computational
speedup. Third, we extend RDS-DeePC to nonlinear systems through a
two-stage online selection framework that combines locality-based
filtering with sensitivity-based robust selection, accelerated by the
LiSSA algorithm \citep{agarwal2017second}.

We validate the approach on four systems of increasing complexity: a
DC motor (LTI), an inverted pendulum, a planar quadrotor UAV tracking
a figure-8 trajectory, and a kinematic bicycle vehicle following a
figure-8 path. Under 20\% data corruption, RDS-DeePC achieves
94–97\% clean segment selection across all systems. On the DC motor,
it yields MSE 0.21 versus 1.09 for fully corrupted data. On the UAV,
it reduces tracking RMSE by 48\% relative to distance-only selection
(0.76 m vs. 1.48 m) while maintaining 97.4\% clean selection.

\section{Preliminaries}
\label{sec:preliminaries}

\subsection{Notation}

We denote by $\R^n$ the $n$-dimensional Euclidean space. For a matrix
$A$, $A^\top$ is its transpose, $A(:,j)$ is its $j$-th column, and
$\|A\|$ is its spectral norm. For a vector $v$ and a positive definite
matrix $P$, $\|v\|_P^2 = v^\top P v$ is the weighted squared norm.
The identity matrix of dimension $n$ is $I_n$, and $e_j$ denotes the
$j$-th standard basis vector. We write $\diag(\cdot)$ for diagonal
matrix construction.

\subsection{System Description}

Consider a discrete-time linear time-invariant (LTI) system:
\begin{equation}
x_{k+1} = Ax_k + Bu_k, \quad y_k = Cx_k + Du_k
\label{eq:lti_system}
\end{equation}
with state $x_k \in \R^n$, input $u_k \in \R^m$, and output
$y_k \in \R^p$ at time step $k$. We assume access to a pre-collected
dataset $\mathcal{D} = \{(u_1, y_1), \ldots, (u_{N_d}, y_{N_d})\}$
of $N_d$ input-output measurements.

\subsection{Hankel Matrix Construction}

From trajectory data of length $N_d$, we construct Hankel matrices
with depth $L = T_{\text{ini}} + N$, where $T_{\text{ini}}$ is the
initial trajectory length used for implicit state estimation and $N$
is the prediction horizon. The input Hankel matrix takes the form:
\begin{equation}
\mathscr{H}_L(u) = \begin{bmatrix}
u_1 & u_2 & \cdots & u_T \\
u_2 & u_3 & \cdots & u_{T+1} \\
\vdots & \vdots & \ddots & \vdots \\
u_L & u_{L+1} & \cdots & u_{N_d}
\end{bmatrix} \in \R^{Lm \times T}
\label{eq:hankel}
\end{equation}
where $T = N_d - L + 1$ is the number of trajectory segments. The
output Hankel matrix $\mathscr{H}_L(y) \in \R^{Lp \times T}$ is
constructed analogously.

We partition the Hankel matrices into past and future components:
\begin{equation}
\begin{bmatrix} U_p \\ U_f \end{bmatrix} = \mathscr{H}_L(u), \quad
\begin{bmatrix} Y_p \\ Y_f \end{bmatrix} = \mathscr{H}_L(y)
\end{equation}
where $U_p \in \R^{T_{\text{ini}} m \times T}$,
$U_f \in \R^{Nm \times T}$,
$Y_p \in \R^{T_{\text{ini}} p \times T}$, and
$Y_f \in \R^{Np \times T}$.

Each column $j \in \{1, \ldots, T\}$ constitutes a
\emph{trajectory segment}:
\begin{equation}
z_j = \begin{bmatrix} U_p(:,j)^\top & U_f(:,j)^\top & Y_p(:,j)^\top
  & Y_f(:,j)^\top \end{bmatrix}^\top
\end{equation}
corresponding to a length-$L$ window of the original data beginning
at time index $j$.

\subsection{Willems' Fundamental Lemma}

DeePC rests on the theoretical foundation of Willems' fundamental
lemma \citep{willems2005note}, which has attracted renewed interest
in the data-driven control community
\citep{markovsky2021behavioral}:

\begin{lem}[Willems et al., 2005]
\label{lem:willems}
Consider an LTI system \eqref{eq:lti_system} of order $n$. If the
input sequence $\{u_1, \ldots, u_{N_d}\}$ is persistently exciting of
order $L + n$, then any valid length-$L$ input-output trajectory
$(\bar{u}, \bar{y})$ of the system can be expressed as:
\begin{equation}
\begin{bmatrix} U_p \\ Y_p \\ U_f \\ Y_f \end{bmatrix} g =
\begin{bmatrix} u_{\text{ini}} \\ y_{\text{ini}} \\ \bar{u} \\
  \bar{y} \end{bmatrix}
\label{eq:willems}
\end{equation}
for some coefficient vector $g \in \R^T$.
\end{lem}

The lemma enables prediction of future trajectories directly from
data, without explicit system identification.

\subsection{Data-Enabled Predictive Control}

Building on Lemma~\ref{lem:willems}, DeePC \citep{coulson2019data}
casts predictive control as a data-driven optimization. In the
presence of noise, the regularized formulation
\citep{coulson2019regularized} solves:
\begin{equation}
\label{eq:deepc_standard}
\begin{aligned}
\min_{g} \quad & \|Y_f g - y_{\text{ref}}\|_Q^2 + \|U_f g\|_R^2
  + \lambda_g \|g\|^2 \\
& + \lambda_y \|Y_p g - y_{\text{ini}}\|^2
  + \lambda_u \|U_p g - u_{\text{ini}}\|^2
\end{aligned}
\end{equation}
where $y_{\text{ref}} \in \R^{Np}$ is the reference trajectory,
$Q \succ 0$ and $R \succ 0$ are tracking and control weights,
$\lambda_g > 0$ regularizes the coefficient vector, and
$\lambda_y, \lambda_u \gg 1$ enforce the initial trajectory
constraints. The first control input $u_0 = (U_f g^*)_{1:m}$ is
applied in a receding horizon fashion.

\subsection{Computational Complexity of Standard DeePC}

Problem~\eqref{eq:deepc_standard} is an unconstrained quadratic
program in $T$ decision variables. The normal equations yield:
\begin{equation}
H g^* = b
\label{eq:normal_equations}
\end{equation}
where $H \in \R^{T \times T}$ is the Hessian and $b \in \R^T$ is the
linear term. Direct solution via Cholesky factorization requires
$O(T^3)$ operations.

For typical applications where $T$ reaches into the thousands, this
cost is too high for real-time MPC.
Table~\ref{tab:complexity_motivation} illustrates the scaling
behavior:

\begin{table}[hb]
\begin{center}
\caption{DeePC computational scaling with dataset
  size}\label{tab:complexity_motivation}
\begin{tabular}{rccc}
\toprule
$T$ (segments) & Operations & Time @ 1 GFLOP/s & Real-time? \\
\midrule
100 & $10^6$ & 1 ms & \checkmark \\
500 & $1.25 \times 10^8$ & 125 ms & Marginal \\
1000 & $10^9$ & 1 s & $\times$ \\
4000 & $6.4 \times 10^{10}$ & 64 s & $\times$ \\
\bottomrule
\end{tabular}
\end{center}
\end{table}

This scaling motivates data selection: a carefully chosen subset of
$K \ll T$ segments reduces complexity to $O(K^3)$, which can enable
real-time control.

\subsection{Influence Functions Background}

Influence functions originate in robust statistics
\citep{hampel1974influence,cook1980characterizations}, where they
quantify how individual data points perturb statistical estimators.
\cite{koh2017understanding} popularized their use in machine learning
for interpreting model predictions, and subsequent work has extended
the theory \citep{bae2022if} and scaled the computation to large
models \citep{schioppa2022scaling,grosse2023studying}.

The core idea proceeds as follows: given an objective function in
which sample $z$'s contribution is scaled by a weight $w$, the
influence function measures how the optimal parameters shift when $w$
is perturbed. This derivative is computed via the implicit function
theorem without re-solving the optimization. Related data valuation
strategies include Data Shapley \citep{ghorbani2019data} and coreset
methods \citep{mirzasoleiman2020coresets}.

\section{Sensitivity Score via Influence Functions}
\label{sec:sensitivity}

We now derive a sensitivity score for each trajectory segment using
influence function analysis. The score quantifies the leverage a given
segment exerts on the DeePC optimization, providing a principled
basis for data selection.

\subsection{Trajectory-Weighted DeePC Formulation}

To isolate individual trajectory contributions, we introduce
per-segment weights $w = [w_1, \ldots, w_T]^\top \in \R^T$ with
$W = \diag(w)$. The weighted DeePC objective becomes:
\begin{equation}
\label{eq:deepc_weighted}
\begin{aligned}
J(g; w) &= \|Y_f W g - y_{\text{ref}}\|_Q^2 + \|U_f W g\|_R^2
  + \lambda_g \|g\|^2 \\
&\quad + \lambda_y \|Y_p W g - y_{\text{ini}}\|^2
  + \lambda_u \|U_p W g - u_{\text{ini}}\|^2
\end{aligned}
\end{equation}

At the baseline $w_j = 1$ for all $j$, this reduces to standard
DeePC \eqref{eq:deepc_standard}. Expanding into quadratic form:
\begin{equation}
J(g; w) = g^\top W M W g - 2 b^\top W g + \lambda_g \|g\|^2
  + \text{const}
\end{equation}
where:
\begin{align}
M &= Y_f^\top Q Y_f + U_f^\top R U_f + \lambda_y Y_p^\top Y_p
  + \lambda_u U_p^\top U_p \label{eq:M_def} \\
b &= Y_f^\top Q y_{\text{ref}} + \lambda_y Y_p^\top y_{\text{ini}}
  + \lambda_u U_p^\top u_{\text{ini}} \label{eq:b_def}
\end{align}

The Hessian at the baseline $w = \mathbf{1}$ is:
\begin{equation}
H = M + \lambda_g I_T
\label{eq:hessian}
\end{equation}

The optimality condition reads:
\begin{equation}
(W M W + \lambda_g I_T) g^*(w) = W b
\label{eq:optimality}
\end{equation}

\subsection{Influence on Optimal Coefficients}

We first characterize how perturbations to segment weights propagate
to the optimal solution.

\begin{prop}[Influence on Optimal Coefficients]
\label{prop:influence_g}
The influence of trajectory segment $z_j$ on the optimal coefficient
vector, evaluated at the baseline $w = \mathbf{1}$, is:
\begin{equation}
\mathcal{I}_g(z_j) := \frac{dg^*}{dw_j}\bigg|_{w=\mathbf{1}}
  = g_j^* H^{-1}(\lambda_g e_j - m_j)
\label{eq:influence_g}
\end{equation}
where $m_j = M(:,j)$ is the $j$-th column of $M$ and $e_j$ is the
$j$-th standard basis vector.
\end{prop}

\begin{pf}
Define the implicit function
$F(g, w) = (WMW + \lambda_g I_T)g - Wb$, satisfying
$F(g^*(w), w) = 0$ at optimality. Using
$\frac{\partial W}{\partial w_j} = e_j e_j^\top$, at baseline
$w = \mathbf{1}$:
\begin{equation}
\frac{\partial F}{\partial w_j}\bigg|_{w=\mathbf{1}}
  = -\lambda_g g_j^* e_j + g_j^* m_j = g_j^*(m_j - \lambda_g e_j)
\end{equation}
where we used the optimality condition $Mg^* = b - \lambda_g g^*$.
By the implicit function theorem:
\begin{equation}
\frac{dg^*}{dw_j} = -H^{-1} \cdot g_j^*(m_j - \lambda_g e_j)
  = g_j^* H^{-1}(\lambda_g e_j - m_j)
\end{equation}
\end{pf}

\subsection{Sensitivity Score Derivation}

To obtain a scalar measure suitable for data selection, we examine
how perturbations affect a test metric. Define the control cost:
\begin{equation}
f(g; w) = \|Y_f W g - y_{\text{ref}}\|_Q^2 + \|U_f W g\|_R^2
\label{eq:test_metric}
\end{equation}

\begin{thm}[Sensitivity Score]
\label{thm:sensitivity}
The sensitivity of trajectory segment $z_j$, defined as the total
derivative of the control cost with respect to segment weight, is:
\begin{equation}
\calS(z_j) := \frac{df(g^*(w); w)}{dw_j}\bigg|_{w=\mathbf{1}}
  = 2\lambda_g g_j^* p_j
\label{eq:sensitivity_score}
\end{equation}
where $p = H^{-1} v$ and
$v = \nabla_g f(g^*; \mathbf{1})
  = 2(Y_f^\top Q(Y_f g^* - y_{\text{ref}}) + U_f^\top R U_f g^*)$.
\end{thm}

\begin{pf}
By the chain rule, the total derivative decomposes into a direct
effect and an indirect effect:
\begin{equation}
\frac{df}{dw_j} = \frac{\partial f}{\partial w_j}\bigg|_{g=g^*}
  + \nabla_g f(g^*)^\top \frac{dg^*}{dw_j}
\label{eq:total_derivative}
\end{equation}

Let $M_f = Y_f^\top Q Y_f + U_f^\top R U_f$ and
$c_f = Y_f^\top Q y_{\text{ref}}$, so that
$v = 2(M_f g^* - c_f)$. The direct effect evaluates to
$g_j^* v_j$. For the indirect effect, set $p = H^{-1} v$ and invoke
$Mp = v - \lambda_g p$ to obtain
$v^\top \mathcal{I}_g(z_j) = g_j^*(2\lambda_g p_j - v_j)$. Summing
both contributions yields $\calS(z_j) = 2\lambda_g g_j^* p_j$.
\end{pf}

\subsection{Interpretation: Sensitivity as Leverage}

The sensitivity score $\calS(z_j) = 2\lambda_g g_j^* p_j$ can be
interpreted as a \emph{leverage measure}: the magnitude
$|\calS(z_j)|$ quantifies how much the control cost would change if
segment $z_j$'s weight were perturbed.

\begin{prop}[High Sensitivity Indicates Outliers]
\label{prop:outlier_interpretation}
When the data pool contains a mixture of clean and corrupted segments,
high-sensitivity segments typically correspond to outliers. Corrupted
segments introduce inconsistencies in the Hankel matrices that
conflict with the bulk of clean data. To accommodate these
inconsistencies, the optimizer tends to assign larger coefficient
magnitudes $|g_j^*|$ to corrupted segments. The inconsistency further
manifests as elevated propagation factors $|p_j|$. Together, both
factors increase $|\calS(z_j)|$ for corrupted data.
\end{prop}

\noindent\textbf{Remark~1} (Connection to Robust Statistics).
\label{rem:robust_stats}
This interpretation aligns with results in robust statistics \citep{hampel1974influence,cook1980characterizations,pan2018robust},
where influence functions identify high-leverage observations.
Algorithmic treatments of adversarial corruption have since been
developed \citep{hopkins2020robust}. In the DeePC setting, trajectory
segments play the role of observations, and $\calS$ quantifies their
leverage on the control optimization.

\subsection{Selection Criterion: Low Sensitivity}

The preceding analysis motivates a simple selection criterion:

\noindent\textbf{Definition~1} (Low-Sensitivity Selection).
\label{def:selection}
Given a selection size $K < T$, the active set is:
\begin{equation}
\calA = \left\{ j : |\calS(z_j)| \text{ is among the } K
  \text{ smallest} \right\}
\label{eq:selection_criterion}
\end{equation}
Equivalently, $\calA = \argmin_{|S|=K} \sum_{j \in S} |\calS(z_j)|$.

This selection achieves outlier filtering, representative data
retention, and computational reduction in one step.

\section{RDS-DeePC Algorithm for LTI Systems}
\label{sec:algorithm}

We now present the complete RDS-DeePC algorithm, comprising an
offline phase for data analysis and selection followed by an online
phase for real-time control.

\subsection{Offline Phase: Sensitivity Analysis and Selection}

\begin{algorithm}[t]
\caption{RDS-DeePC: Offline Phase}
\label{alg:offline}
\begin{algorithmic}[1]
\Require Hankel matrices $U_p, U_f, Y_p, Y_f \in \R^{\cdot \times T}$
\Require Parameters $Q, R, \lambda_g, \lambda_y, \lambda_u$;
  selection size $K$
\Require Nominal $(u_{\text{ini}}, y_{\text{ini}})$, reference
  $y_{\text{ref}}$
\State Construct $M$ via \eqref{eq:M_def}, $H \leftarrow M + \lambda_g I_T$,
  $b$ via \eqref{eq:b_def}
\State $g^* \leftarrow H^{-1} b$ (Cholesky factorization)
\State $M_f \leftarrow Y_f^\top Q Y_f + U_f^\top R U_f$,
  $c_f \leftarrow Y_f^\top Q y_{\text{ref}}$
\State $v \leftarrow 2(M_f g^* - c_f)$,
  $p \leftarrow H^{-1} v$ (reuse Cholesky)
\For{$j = 1, \ldots, T$}
    \State $\calS(z_j) \leftarrow 2\lambda_g g_j^* p_j$
\EndFor
\State $\calA \leftarrow \text{argsort}(|\calS|)_{1:K}$
  ($K$ smallest $|\calS|$)
\State Construct reduced matrices indexed by $\calA$;
  compute/store Cholesky of $\tilde{H}$
\State \Return Reduced matrices, Cholesky factor, active set $\calA$
\end{algorithmic}
\end{algorithm}

\subsection{Online Phase: Real-Time Control}

\begin{algorithm}[t]
\caption{RDS-DeePC: Online Phase}
\label{alg:online}
\begin{algorithmic}[1]
\Require Reduced matrices
  $\tilde{U}_p, \tilde{U}_f, \tilde{Y}_p, \tilde{Y}_f
  \in \R^{\cdot \times K}$
\Require Precomputed Cholesky factor of $\tilde{H}$
\Require Parameters $Q, R, \lambda_g, \lambda_y, \lambda_u$
\For{each time step $t = 0, 1, 2, \ldots$}
    \State Measure current state, construct
      $(u_{\text{ini}}, y_{\text{ini}})$
    \State $\tilde{b} \leftarrow \tilde{Y}_f^\top Q y_{\text{ref}}
      + \lambda_y \tilde{Y}_p^\top y_{\text{ini}}
      + \lambda_u \tilde{U}_p^\top u_{\text{ini}}$
    \State $\tilde{g}^* \leftarrow \tilde{H}^{-1} \tilde{b}$
      (using stored Cholesky factor)
    \State Apply $u_t \leftarrow (\tilde{U}_f \tilde{g}^*)_{1:m}$
\EndFor
\end{algorithmic}
\end{algorithm}

\subsection{Computational Complexity Analysis}

\begin{prop}[Complexity Analysis]
\label{prop:complexity}
Let $T$ denote the total number of trajectory segments and $K$ the
selection size. The offline phase costs $O(T^3)$ plus $O(T)$ for
sensitivity computation. The online phase costs $O(K^2)$ per MPC step
using the precomputed Cholesky factor. The resulting online speedup
over full DeePC is $(T/K)^2$ for iterative solvers or $(T/K)^3$ for
direct solvers.
\end{prop}

\subsection{Selection Size Guidelines}

The selection size $K$ trades off robustness against information
retention. A range of $K \in [30, 100]$ typically provides robust
performance with large speedup. As a practical heuristic,
$K \approx 5$--$10\times$ the effective system order serves well.

\section{Extension to Nonlinear Systems}
\label{sec:nonlinear}

The RDS-DeePC algorithm of Section~\ref{sec:algorithm} assumes LTI
dynamics, yet many practical applications involve nonlinear systems.
This section extends the framework to the nonlinear setting through a
two-stage online selection scheme.

\subsection{Challenges for Nonlinear Systems}

For a nonlinear system $\dot{x} = f(x, u)$, Willems' fundamental
lemma does not hold globally. Near a given operating trajectory,
however, the system behaves approximately linearly. Two consequences
follow. First, data from distant operating regions should be
excluded, since they reflect dynamics that may differ from those
at the current operating point (\emph{local relevance}). Second, the
sensitivity score depends on the current operating point and must
therefore be recomputed online (\emph{dynamic sensitivity}).

\subsection{Two-Stage Selection Framework}

We propose a two-stage framework: an initial filter based on
locality, followed by robust selection within the resulting local
subset.

\subsubsection{Stage 1: Locality-Based Filtering.}
Given the current initial trajectory $(u_{\text{ini}}, y_{\text{ini}})$,
define a distance metric:
\begin{equation}
d(z_j, z_t) = \left\| \begin{bmatrix} Y_p(:,j) \\ U_p(:,j)
  \end{bmatrix} - \begin{bmatrix} y_{\text{ini}} \\
  u_{\text{ini}} \end{bmatrix} \right\|_W
\label{eq:distance}
\end{equation}
where $W \succ 0$ is a weighting matrix. The local subset $\calL_t$
retains the $K_L$ closest segments.

\subsubsection{Stage 2: Sensitivity-Based Robust Selection.}
Within $\calL_t$, we compute local sensitivity scores and retain the
$K_R$ segments exhibiting the lowest sensitivity:
\begin{equation}
\calA_t = \left\{ j \in \calL_t : |\calS^{(\calL_t)}(z_j)|
  \text{ among } K_R \text{ smallest} \right\}
\end{equation}

\subsection{Efficient Computation via LiSSA}

Computing sensitivity scores requires the solve $p = H^{-1} v$. We
employ the Linear time Stochastic Second-order Algorithm (LiSSA)
\citep{agarwal2017second} for efficient approximation, exploiting
the Neumann series expansion:
\begin{equation}
\tilde{p}^{(i+1)} = v + (I - \alpha H) \tilde{p}^{(i)}, \quad
  \tilde{p}^{(0)} = v
\label{eq:lissa}
\end{equation}

\begin{prop}[LiSSA Complexity]
\label{prop:lissa_complexity}
LiSSA requires $O(K_L \cdot r \cdot s)$ operations to approximate
$H^{-1}v$, where $r$ is the recursion depth and $s$ the sample count.
For typical values $r = 50$, $s = 5$, this cost is well below
direct $O(K_L^3)$ inversion whenever $K_L > 250$.
\end{prop}

\subsection{Online RDS-DeePC Algorithm}

\begin{algorithm}[t]
\caption{Online RDS-DeePC for Nonlinear Systems}
\label{alg:online_nonlinear}
\begin{algorithmic}[1]
\Require Data pool $\{z_1, \ldots, z_T\}$;
  parameters $Q, R, \lambda_g, \lambda_y, \lambda_u$
\Require Selection sizes $K_L$ (local), $K_R$ (robust);
  LiSSA parameters $\alpha, r, s$
\For{each time step $t$}
    \State Measure $(u_{\text{ini}}, y_{\text{ini}})$;
      compute distances $d_j$ via \eqref{eq:distance}
    \State $\calL_t \leftarrow \text{argsort}(d)_{1:K_L}$
    \State Construct local matrices; solve
      $g_L^* = H_L^{-1} b_L$
    \State Compute $v_L$; approximate
      $p_L \approx H_L^{-1} v_L$ via LiSSA \eqref{eq:lissa}
    \For{$j \in \calL_t$}
        \State $\calS_j \leftarrow 2\lambda_g (g_L^*)_j (p_L)_j$
    \EndFor
    \State $\calA_t \leftarrow \text{argsort}(|\calS|)_{1:K_R}$
      within $\calL_t$
    \State Solve reduced DeePC on $\calA_t$;
      apply $u_t \leftarrow (\tilde{U}_f \tilde{g}^*)_{1:m}$
\EndFor
\end{algorithmic}
\end{algorithm}

\subsection{Computational Complexity}

\begin{prop}[Online Complexity]
\label{prop:online_complexity}
The per-step complexity of Algorithm~\ref{alg:online_nonlinear} is
$O(T) + O(K_L \cdot r \cdot s) + O(K_R^2)$, reflecting distance
computation, LiSSA-based sensitivity estimation, and the reduced
DeePC solve, respectively.
\end{prop}

For typical parameters ($T = 4050$, $K_L = 200$, $K_R = 30$,
$r = 50$, $s = 5$), the dominant cost amounts to roughly 50,000
operations per step—well within the budget for real-time control at
sampling rates of up to several hundred Hertz.

\section{Numerical Results}
\label{sec:results}

We evaluate RDS-DeePC on four systems of increasing complexity, each
subject to 20\% corrupted data. Three data selection methods are
compared throughout: \textbf{RDS-DeePC} (two-stage locality +
sensitivity selection), \textbf{Distance} (locality-based selection
alone), and \textbf{Random} (uniform random selection). For nonlinear
systems, the control architecture pairs a baseline controller (LQR or
pure pursuit) with DeePC corrections clipped to the persistent
excitation training range.

\subsection{DC Motor Position Control}

We first consider a DC motor with dynamics
$J\dot{\omega} = K_t i_a - b\omega$,
$L_a \dot{i}_a = V_a - R_a i_a - K_e \omega$, discretized at 50 ms.
The dataset comprises 50 trajectories ($T = 4050$ segments), of which
20\% are corrupted via high noise ($15\times$ nominal), sensor bias
($\pm 0.5$ rad), and input-output mismatch.

\begin{table}[hb]
\begin{center}
\caption{DC motor performance with 20\% corrupted
  data}\label{tab:corrupt_results}
\begin{tabular}{lcccc}
\toprule
Method & $K$ & MSE [rad$^2$] & RMSE [$^\circ$] & Clean/$K$ \\
\midrule
Clean-only & 4050 & 0.0790 & 16.10 & all \\
Full (w/ corrupt) & 4050 & 0.7955 & 51.10 & 3240/4050 \\
\midrule
RDS-DeePC & 30 & \textbf{0.0790} & 16.10 & 29/30 \\
RDS-DeePC & 60 & \textbf{0.0791} & 16.12 & 56/60 \\
RDS-DeePC & 90 & \textbf{0.0790} & 16.11 & 84/90 \\
\midrule
Random & 30 & 13.93 $\pm$ 23.77 & 213.87 & $\sim$23/30 \\
Random & 60 & 13.25 $\pm$ 6.94 & 208.52 & $\sim$47/60 \\
Random & 90 & 6.93 $\pm$ 6.47 & 150.78 & $\sim$71/90 \\
\bottomrule
\end{tabular}
\end{center}
\end{table}

\begin{figure}
\begin{center}
\includegraphics[width=8.4cm]{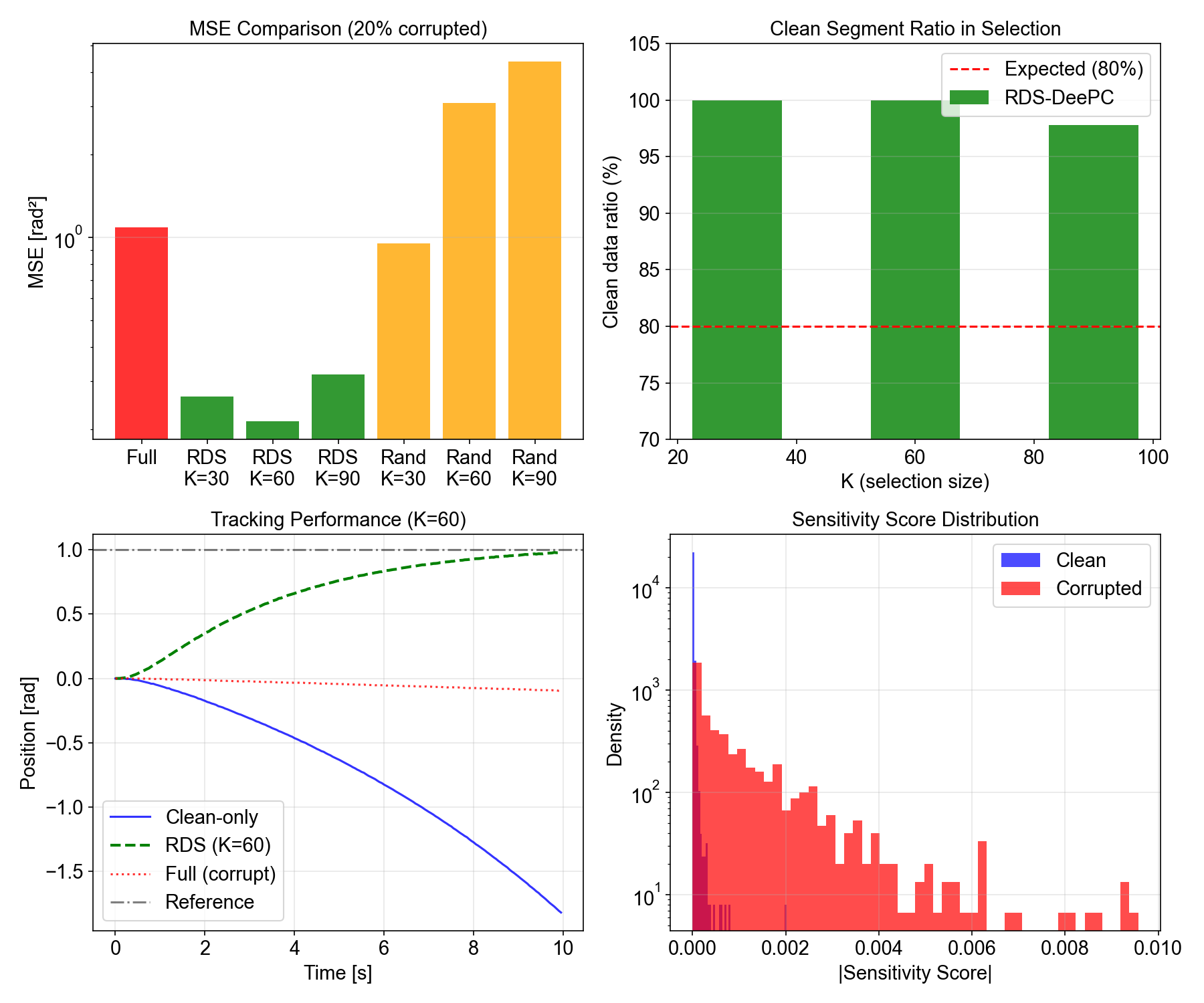}
\caption{DC motor with 20\% corrupted data. Top-left: MSE comparison.
  Top-right: Clean data ratio (RDS-DeePC: 93--97\% vs.\ expected
  80\%). Bottom-left: Tracking at $K=60$. Bottom-right: Influence
  score separation.}
\label{fig:dcmotor_results}
\end{center}
\end{figure}

As Table~\ref{tab:corrupt_results} shows, corruption degrades full
DeePC from MSE 0.079 to 0.796~rad$^2$. Random selection fails with
high variance (MSE $13.93 \pm 23.77$ at $K=30$). RDS-DeePC, by
contrast, recovers clean-data performance across all values of $K$,
automatically selecting 93--97\% clean segments
(Fig.~\ref{fig:dcmotor_results}). The improvement over random
selection reaches 99.4\%.

\subsection{Inverted Pendulum (Nonlinear)}

For nonlinear validation, we consider cart-pole stabilization from
$\theta_0 = 30^\circ$ to the upright equilibrium. The dynamics follow
$(m_c + m_p)\ddot{x}_c + m_p l \ddot{\theta} \cos\theta
  = F + m_p l \dot{\theta}^2 \sin\theta$
with $m_c = 1$~kg, $m_p = 0.1$~kg, $l = 0.5$~m. We collect 100
trajectories with 20\% corrupted and set $K_L = 200$ for locality
filtering and $K_R = 30$ for sensitivity selection.

\begin{figure}
\begin{center}
\includegraphics[width=8.4cm]{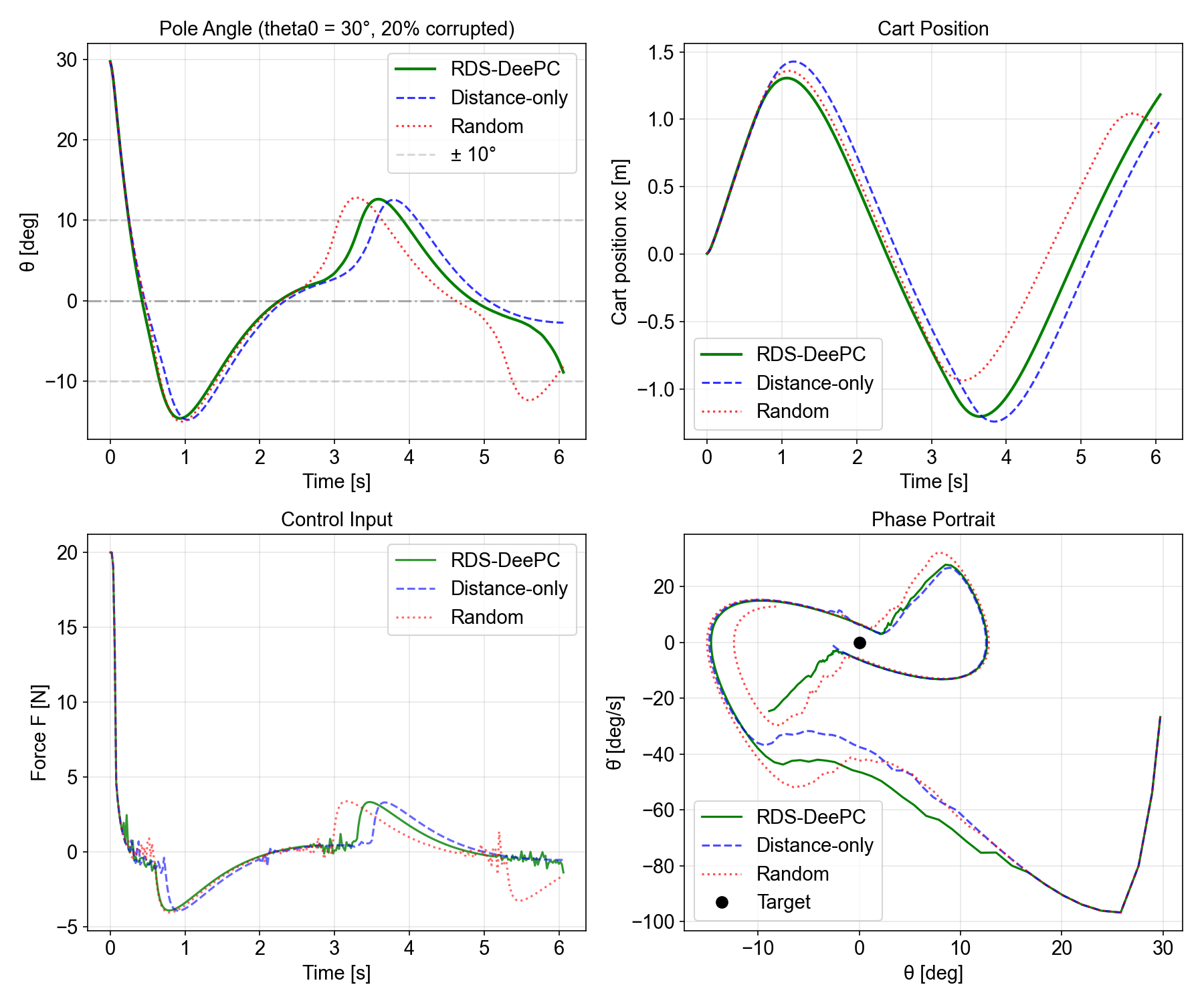}
\caption{Inverted pendulum ($\theta_0 = 30^\circ$, 20\% corrupted).
  Top-left: RDS-DeePC stabilizes within $\pm 10^\circ$; Random fails.
  Top-right: Cart position. Bottom-left: Control input. Bottom-right:
  Phase portrait.}
\label{fig:pendulum_results}
\end{center}
\end{figure}

\begin{table}[hb]
\begin{center}
\caption{Inverted pendulum: stabilization
  metrics}\label{tab:pendulum_results}
\begin{tabular}{lccc}
\toprule
Method & Settling [s] & Max $|\theta|$ [$^\circ$]
  & Final $|x_c|$ [m] \\
\midrule
RDS-DeePC & \textbf{3.9} & 29.7 & 1.18 \\
Distance & 4.1 & 29.7 & 0.99 \\
Random & 5.9 & 29.7 & 0.89 \\
\bottomrule
\end{tabular}
\end{center}
\end{table}

Fig.~\ref{fig:pendulum_results} and
Table~\ref{tab:pendulum_results} confirm that RDS-DeePC achieves the
fastest settling time (3.9~s versus 4.1~s for Distance and 5.9~s for
Random---a 34\% improvement over random selection). The advantage of
two-stage selection is clearest during the transient phase,
where sensitivity-based filtering removes corrupted segments that
would otherwise introduce destabilizing corrections.

\subsection{Planar Quadrotor UAV (Figure-8 Tracking)}

To test the method on a higher-dimensional nonlinear system, we
consider a planar quadrotor with Crazyflie-class parameters
($m = 0.027$~kg, $I_{xx} = 1.4 \times 10^{-5}$~kg\,m$^2$, arm
length $d = 0.0397$~m), discretized at $\Delta t = 0.02$~s. The
6-state model comprises $(p_x, v_x, p_z, v_z, \phi, \dot{\phi})$,
with inputs (left/right thrust $f_1, f_2$) and outputs
$(p_x, p_z, \phi)$.

The task is to track a figure-8 (lemniscate) trajectory in the
$p_x$--$p_z$ plane:
\begin{equation}
p_x(t) = r_x \sin(\omega t), \quad
p_z(t) = z_{\text{off}} + r_z \sin(2\omega t)
\end{equation}
with $r_x = 1.0$~m, $r_z = 0.5$~m, $z_{\text{off}} = 1.0$~m, and
period $= 4.0$~s. A continuous-time LQR controller provides baseline
tracking, and DeePC corrections are clipped to $\pm 0.04$~N. We
collect 150 trajectories (200 steps each) with 20\% corrupted
($T = 28050$ segments) and set $K_L = 600$, $K_R = 60$.

\begin{table}[hb]
\begin{center}
\caption{Planar quadrotor: figure-8 tracking over
  3 loops}\label{tab:uav_results}
\begin{tabular}{lccc}
\toprule
Method & RMSE [m] & MaxErr [m] & Max$|\phi|$ [$^\circ$] \\
\midrule
RDS-DeePC & \textbf{0.7638} & \textbf{1.3671} & 49.2 \\
Random & 1.0266 & 1.9752 & 24.7 \\
Distance & 1.4827 & 2.8521 & 59.6 \\
\midrule
\multicolumn{4}{l}{\small Clean selection: RDS 97.4\%, Distance 67.0\%} \\
\bottomrule
\end{tabular}
\end{center}
\end{table}

\begin{figure}
\begin{center}
\includegraphics[width=8.4cm]{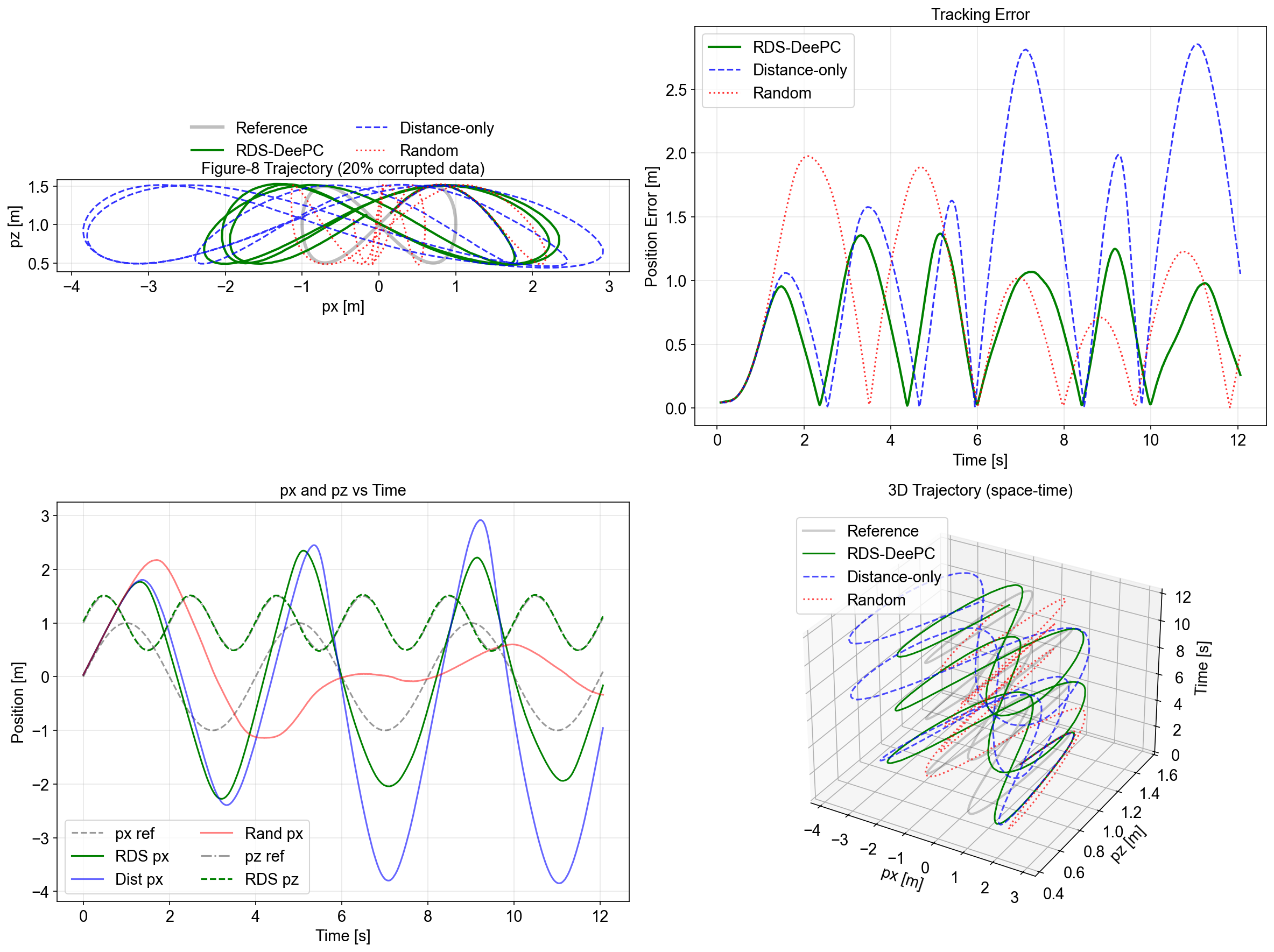}
\caption{Planar quadrotor figure-8 tracking (20\% corrupted). Top-left:
  trajectory in $p_x$--$p_z$ plane. Top-right: position tracking
  error over time. Bottom-left: $p_x$ and $p_z$ vs.\ time.
  Bottom-right: 3D space--time view.}
\label{fig:uav_results}
\end{center}
\end{figure}

Table~\ref{tab:uav_results} and Fig.~\ref{fig:uav_results}
demonstrate that RDS-DeePC achieves the lowest RMSE (0.76~m),
outperforming distance-only selection by 48\% and random selection by
26\%. The clean selection rate—97.4\% for RDS-DeePC versus just
67.0\% for distance-based selection—reveals that proximity alone is
an insufficient filter: many nearby segments are corrupted, and only
sensitivity-based scoring can distinguish them. This result shows
the main advantage of two-stage selection: when corruption is
spatially uncorrelated with the operating trajectory, the sensitivity
score provides filtering capability that locality alone lacks.

\subsection{Kinematic Bicycle Vehicle (Figure-8 Path Tracking)}

We further evaluate a kinematic bicycle vehicle with a wheelbase
of $L = 2.7$~m, discretized at $\Delta t = 0.05$~s. The 4-state model
has states $(X, Y, \psi, v)$, inputs (steering angle $\delta$,
acceleration $a$), and outputs $(X, Y, \psi)$.

The task is to track a figure-8 (lemniscate of Gerono) path at
$v_{\text{ref}} = 5.0$~m/s:
\begin{equation}
X(t) = r_x \sin(t), \quad Y(t) = \frac{r_y}{2} \sin(2t)
\end{equation}
with $r_x = 30.0$~m, $r_y = 15.0$~m (perimeter $\approx 142.6$~m).
A pure-pursuit controller provides baseline tracking, and DeePC
corrections are clipped to $\pm 0.10$~rad (steering) and
$\pm 1.0$~m/s$^2$ (acceleration). We collect 120 trajectories
(150 steps each) with 20\% corrupted ($T = 16440$ segments) and set
$K_L = 500$, $K_R = 50$.

\begin{table}[hb]
\begin{center}
\caption{Bicycle vehicle: figure-8 path tracking over
  2 laps}\label{tab:vehicle_results}
\begin{tabular}{lccc}
\toprule
Method & RMSE [m] & MaxLat [m] & Mean$|\Delta\psi|$ [$^\circ$] \\
\midrule
Distance & \textbf{0.1841} & \textbf{0.5507} & \textbf{2.29} \\
RDS-DeePC & 0.2373 & 0.5652 & 4.12 \\
Random & 0.6464 & 1.3847 & 5.74 \\
\midrule
\multicolumn{4}{l}{\small Clean selection: RDS 94.7\%, Distance 97.7\%} \\
\bottomrule
\end{tabular}
\end{center}
\end{table}

\begin{figure}
\begin{center}
\includegraphics[width=8.4cm]{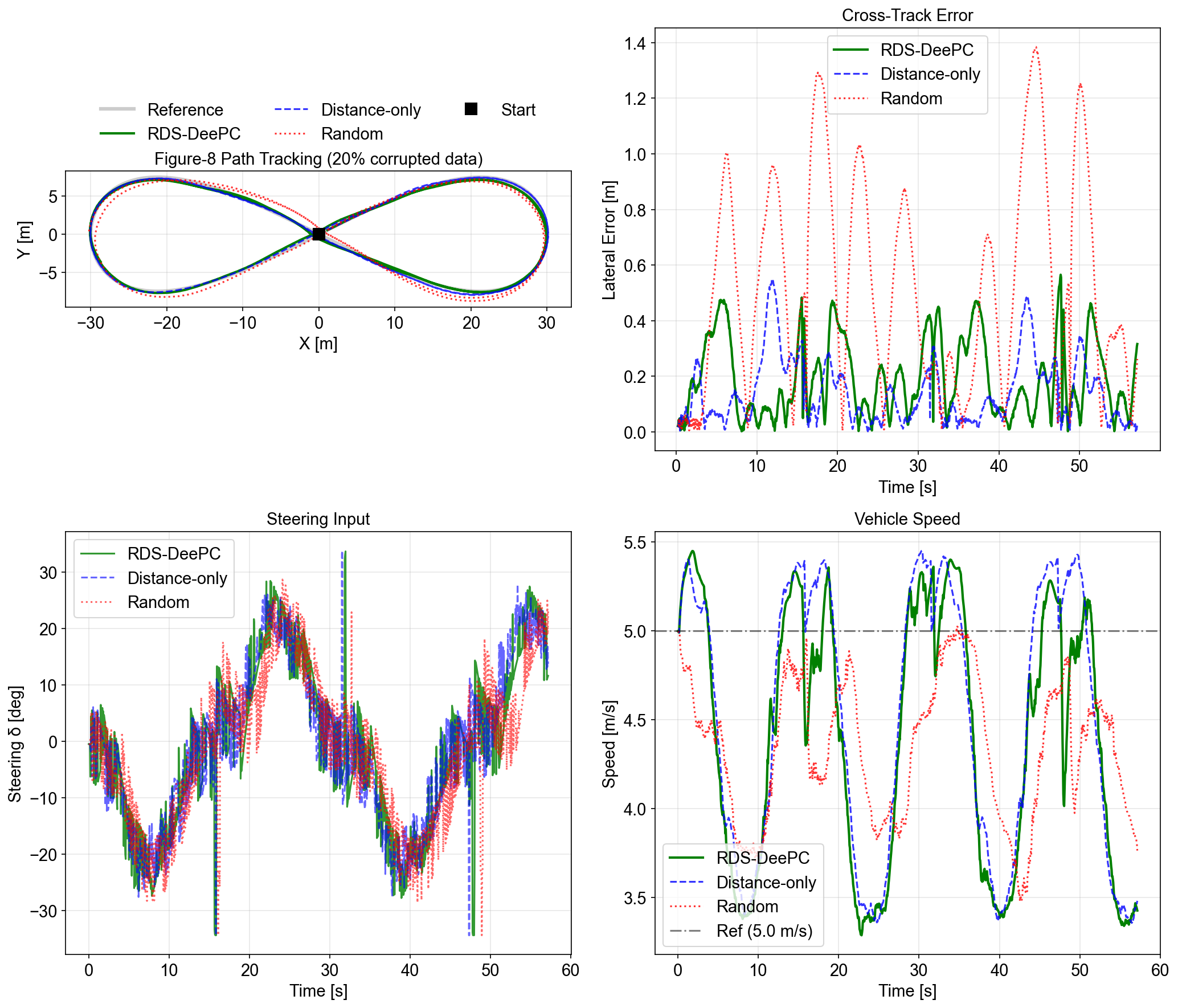}
\caption{Bicycle vehicle figure-8 tracking (20\% corrupted). Top-left:
  path in $X$--$Y$ plane. Top-right: lateral tracking error.
  Bottom-left: steering angle $\delta$. Bottom-right: longitudinal
  speed $v$.}
\label{fig:vehicle_results}
\end{center}
\end{figure}

Table~\ref{tab:vehicle_results} and Fig.~\ref{fig:vehicle_results}
show that distance-based selection achieves the lowest RMSE
(0.18~m), slightly outperforming RDS-DeePC (0.24~m). This outcome
is expected: the kinematic bicycle model's dynamics vary smoothly
with the operating point, and the pure-pursuit baseline already provides
accurate nominal tracking. When corruption is spatially correlated
with distance from the operating trajectory---here, distance-based
selection reaches 97.7\% clean data---the additional sensitivity
computation adds only marginal benefit. Both RDS-DeePC and Distance
still outperform Random (0.65~m) by a wide margin, confirming that
data selection remains necessary.

\subsection{Discussion}

Taken together, the four experiments reveal a clear pattern.
Sensitivity-based selection is most valuable when two conditions
coincide: the system is sensitive to data quality, and
locality alone cannot filter corrupted segments. The UAV
experiment illustrates this—distance--based selection
achieves only 67\% clean rate, whereas RDS-DeePC reaches 97.4\%,
yielding a 48\% RMSE improvement. In the vehicle experiment, by
contrast, corruption correlates spatially with distance, so locality
filtering already achieves 97.7\% clean selection, and the sensitivity
stage adds little.

These results suggest a practical guideline: deploy two-stage
RDS-DeePC when the corruption mechanism is independent of the operating
point; consider distance-only selection when the computational budget is
tight and corruption correlates with distance.

\section{Conclusion}
\label{sec:conclusion}

This paper introduces RDS-DeePC, a robust data selection framework
for Data-Enabled Predictive Control that addresses both computational
cost and sensitivity to data quality through influence function
analysis. The main component is the sensitivity score
$\calS(z_j) = 2\lambda_g g_j^* p_j$, which identifies high-leverage
outliers while retaining consistent, low-sensitivity data.

Experiments on four systems of increasing complexity validated the
approach. On the LTI DC motor, RDS-DeePC reduces MSE from 1.09
(fully corrupted data) to 0.21 while selecting only clean segments.
On the nonlinear inverted pendulum, it achieves 34\% faster settling
than random selection. On the planar quadrotor UAV tracking a
figure-8 trajectory, RDS-DeePC improves RMSE by 48\% over
distance-only selection (0.76~m vs.\ 1.48~m) with 97.4\% clean data
selection, showing the importance of sensitivity-based filtering when
locality alone is not enough. On the kinematic bicycle vehicle,
distance-based selection is competitive (0.18~m vs.\ 0.24~m for
RDS-DeePC), indicating that when corruption correlates with spatial
distance, locality-based filtering is sufficient.

These results suggest a practical guideline: use two-stage RDS-DeePC when corruption is spatially uncorrelated with the
operating trajectory; use distance-only selection when corruption
correlates with distance and computational resources are limited. Future
work will explore the adaptive determination of the selection size,
theoretical performance guaranties, and extensions to constrained MPC
formulations.

\section*{DECLARATION OF GENERATIVE AI AND AI-ASSISTED TECHNOLOGIES
  IN THE WRITING PROCESS}

During the preparation of this work, the author used Claude to assist with the language editing of the manuscript. After using this tool, the author reviewed and edited the content as needed and takes full responsibility for the content of the manuscript. 


\bibliography{ifacconf}

\appendix
\section{Proof Details for Theorem~\ref{thm:sensitivity}}
\label{app:proof}

The weighted test metric is
$f(g; w) = g^\top W M_f W g - 2 c_f^\top W g + \text{const}$,
where $M_f = Y_f^\top Q Y_f + U_f^\top R U_f$ and
$c_f = Y_f^\top Q y_{\text{ref}}$.

\emph{Direct effect.} At baseline $w = \mathbf{1}$ and $g = g^*$:
$\frac{\partial f}{\partial w_j} = g_j^* v_j$,
where $v = 2(M_f g^* - c_f)$.

\emph{Indirect effect.} From Proposition~\ref{prop:influence_g},
with $p = H^{-1} v$ and using $Mp = v - \lambda_g p$:
$v^\top \mathcal{I}_g(z_j) = g_j^*(2\lambda_g p_j - v_j)$.

\emph{Combined.}
$\calS(z_j) = g_j^* v_j + g_j^*(2\lambda_g p_j - v_j)
  = 2\lambda_g g_j^* p_j$. \hfill $\blacksquare$

\section{LiSSA Convergence Analysis}
\label{app:lissa}

\begin{prop}[LiSSA Convergence]
Let $\kappa = \lambda_{\max}(H) / \lambda_{\min}(H)$ be the condition
number. With scale $\alpha = 1/\lambda_{\max}(H)$, after $r$
iterations:
\begin{equation}
\|\tilde{p}^{(r)} - H^{-1}v\| \leq \left(1 - \frac{1}{\kappa}
  \right)^r \|H^{-1}v\|
\end{equation}
\end{prop}

\begin{pf}
The iteration $\tilde{p}^{(i+1)} = v + (I - \alpha H)\tilde{p}^{(i)}$
has a fixed point $p^* = H^{-1}v$ and satisfies
$\tilde{p}^{(i+1)} - p^* = (I - \alpha H)(\tilde{p}^{(i)} - p^*)$.
With $\alpha = 1/\lambda_{\max}(H)$, the spectral radius of
$I - \alpha H$ is $\rho(I - \alpha H) = 1 - 1/\kappa$.
The result follows from contraction. For DeePC with regularization
$\lambda_g > 0$, the Hessian $H = M + \lambda_g I$ is
well-conditioned, ensuring rapid convergence.
\end{pf}

\end{document}